\documentclass[twocolumn,showpacs,preprintnumbers,amsmath,amssymb]{revtex4}

\usepackage{graphicx}
\usepackage{dcolumn}
\usepackage{bm}

\begin{document}

\preprint{APS/123-QED}

\title{New pathway to bypass the $^{15}$O waiting point}

\author{I. Stefan$^{1,2}$, F. de Oliveira Santos$^{1}$, M.G. Pellegriti$^1$, G. Dumitru$^2$, J.C. Ang\'{e}lique$^3$, M. Ang\'{e}lique$^1$, E.
Berthoumieux$^4$, A. Buta$^2$, R. Borcea$^2$, A. Coc$^5$, J.M.
Daugas$^6$, T. Davinson$^7$, M. Fadil$^1$, S. Gr\'{e}vy$^1$, J.
Kiener$^5$, A. Lefebvre-Schuhl$^5$, M. Lenhardt$^1$, M.
Lewitowicz$^1$, F. Negoita$^2$, D. Pantelica$^2$, L. Perrot$^1$, O.
Roig$^6$, M.G. Saint Laurent$^1$, I. Ray$^1$, O. Sorlin$^1$, M.
Stanoiu$^8$, C. Stodel$^1$, V. Tatischeff$^3$, J.C. Thomas$^1$}

\address{
$^1$ Grand Acc\'el\'erateur National d'Ions Lourds B.P. 5027 F-14076 Caen Cedex, France \\
$^2$ Horia Hulubei National Institute of Physics and Nuclear Engineering P.O. Box MG6 Bucharest-Margurele, Romania \\
$^3$ Laboratoire de Physique Corpusculaire IN2P3-CNRS ISMRA et Universit\'e de Caen F-14050 Caen, France \\
$^4$ CEA Saclay DSM/DAPNIA/SPHN F-91191 Gif-sur-Yvette, France \\
$^5$ CSNSM CNRS/IN2P3/UPS B\^at.~104 91405 Orsay Campus, France \\
$^6$ CEA/DIF/DPTA/PN BP 12 91680 Bruy\`{e}res le Ch\^{a}tel, France\\
$^7$ Department of Physics and Astronomy University of Edinburgh
Edinburgh EH9 3JZ, United Kingdom \\
$^8$ Institut de Physique Nucl\'eaire IN2P3-CNRS F-91406 Orsay, France \\
}

\date{\today}

\begin{abstract}
We propose the sequential reaction process
$^{15}$O($p$,$\gamma)(\beta^{+}$)$^{16}$O as a new pathway to bypass
of the $^{15}$O waiting point. This exotic reaction is found to have
a surprisingly high cross section, approximately 10$^{10}$ times
higher than the $^{15}$O($p$,$\beta^{+}$)$^{16}$O. These cross
sections were calculated after precise measurements of energies and
widths of the proton-unbound $^{16}$F low lying states, obtained
using the H($^{15}$O,p)$^{15}$O reaction. The large
$(p,\gamma)(\beta^{+})$ cross section can be understood to arise
from the more efficient feeding of the low energy wing of the ground
state resonance by the gamma decay. The implications of the new
reaction in novae explosions and X-ray bursts are discussed.
\end{abstract}

\maketitle

Unbound nuclei play a major role in astrophysics. The proton-unbound
$^{2}$He and the alpha-unbound $^{8}$Be nuclei illustrate this fact.
The former is involved in the p($p$,$\beta^{+}$)d reaction, first
reaction of the $pp$ chain of reactions governing the energy
generation in the sun \cite{1,2}. The latter, whose lifetime is
about 10$^{-16}$ seconds, is involved in the triple alpha reaction
\cite{2,3} which is at the origin of the formation of all the
heavier elements. The proton-unbound nuclei $^{15}$F and $^{16}$F
play an important role in X-ray bursts. These astronomical events
are known to happen in close binary systems, where accretion takes
place from an extended companion star on the surface of a neutron
star (type I X-ray burst). The accreted matter is compressed until
it reaches sufficiently high pressure conditions to trigger a
thermonuclear runaway. In these explosive events, the carbon and
nitrogen elements are mainly transformed into $^{14}$O and $^{15}$O
by successive proton captures \cite{4,5}. Then, the pathway for new
proton captures is hindered by the proton-unbound nuclei $^{15}$F
and $^{16}$F. The reaction flux and the energy generation are then
limited by the relatively slow $\beta^{+}$-decay of $^{14}$O
(t$_{1/2}$=71 s) and $^{15}$O (t$_{1/2}$=122 s), which create
waiting points. The sudden and intense release of energy observed in
X-ray bursts requires to circumvent the limited energy generation in
breakout reactions. The $^{15}$O($\alpha$,$\gamma$)$^{19}$Ne
reaction is considered to be one of the key reactions in this
context \cite{4,5}. It makes the transition into the nucleosynthetic
$rp$ process (rapid proton capture) which is responsible for an
increased rate of energy generation and the synthesis of heavier
elements \cite{6}. In such explosive environments, $^{16}$F is
strongly populated in the ground state (g.s.) or in the first
excited state, and leads to an equilibrium between formation and
decay of this proton-unbound nucleus. From time to time before the
proton is emitted, $^{16}$F can capture another proton thus
producing the $^{17}$Ne particle stable isotope \cite{7}. This
two-proton capture process was calculated to be significant for
extreme densities (larger than 10$^{11}$ g/cm$^{3}$). In this
letter, $\beta^{+}$-decay of $^{16}$F to $^{16}$O is proposed as an
alternative channel. Two reactions $^{15}$O($p$,$\beta^{+}$)$^{16}$O
and $^{15}$O($p$,$\gamma)(\beta^{+}$)$^{16}$O are studied. Both
reactions eventually proceed through the $\beta^{+}$-decay of the
intermediate unbound $^{16}$F g.s., which is fed directly by a
proton capture or indirectly through a proton capture to the first
excited state followed by a $\gamma$-emission. The $\gamma$-decay
occurs noticeably to the low energy wing of the g.s. resonance.
Subsequent proton emission is dramatically hindered due to the fact
that the low energy proton has to tunnel through the Coulomb
potential of the $^{15}$O nucleus. These reaction channels have not
been investigated so far and could speed-up the energy generation,
competing with breakout reactions. The calculation of these reaction
cross sections require the measurement of the energies, widths,
spins and parities of the low lying states of $^{16}$F. These were
obtained from the measurement of the H($^{15}$O,$p$)$^{15}$O
resonant elastic excitation function using low energy $^{15}$O beam
at the SPIRAL facility.

The beam of radioactive $^{15}$O nuclei was produced at the
SPIRAL-GANIL facility through the projectile fragmentation of a 95
A.MeV $^{16}$O primary beam on a thick carbon target. Mean
intensities of 10$^{7}$ pps at an energy of 1.2 A.MeV were obtained
after post acceleration by the CIME cyclotron. A beam contamination
of less than 1 $\%$ of $^{15}$N was achieved using a vertical
betatron oscillation selection device \cite{8} and a suitable
degrader in the analysis line of LISE spectrometer \cite{9} where
the measurements were made. Two stable beams, $^{14}$N and $^{15}$N,
were also used in similar experimental conditions for calibrations.
The excitation function for the elastic scattering at these low
energies can be described by the Rutherford scattering, but shows
"anomalies", i.e. various resonances that are related to individual
states in the compound nucleus. The principle of the measurement is
described in \cite{10,11} and references therein. A 31(1) $\mu$m
thick polyethylene (CH$_{2}$)$_{n}$ target was used, thick enough to
stop the beam inside. The scattered protons were detected by a
silicon detector, placed at forward angles (180$^{\circ}$ in the
center of mass frame) within an angular acceptance of 2$^{\circ}$.
Protons were identified using their energy and time-of-flight. The
energy resolution was 3 keV in the center of mass frame. Fig. 1
shows the excitation function for the H($^{15}$O,$p$)$^{15}$O
reaction measured from $0.450$ MeV to 1.1 MeV. The measured cross
section was reproduced by an R-matrix \cite{12} calculation using
the code ANARKI \cite{13} which is seen to be in a good agreement
with the data. A value of S$_{p}$ = -534 $\pm$ 5 keV was obtained
for the proton separation energy in agreement with the recommended
value \cite{14}. The R-matrix analysis was also used to extract the
properties of the first three states in $^{16}$F, given in Table I.
A significant difference was found between the present and the
recommended value of the width for the first excited state
\cite{14}. This width is an important parameter used in the
calculations presented in the next section.

\begin{figure}
\resizebox{0.48\textwidth}{!}{%
\includegraphics{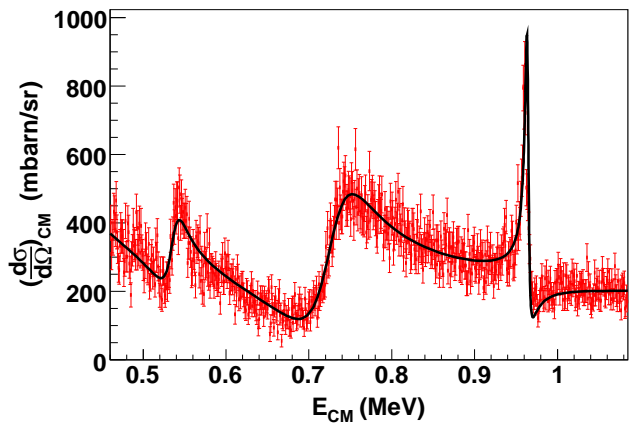}
} \caption{\label{fig:n15}Excitation function for the
H($^{15}$O,$p$)$^{15}$O reaction at 180$^{\circ}$ in the center of
mass frame. The line is a result of an R-matrix calculation using
parameters from Table I.}
\end{figure}
\begin{table}
\caption{\label{tab:table3}Measured properties for the low-lying
states in $^{16}$F.}
\begin{ruledtabular}
\begin{tabular}{cccccc}
E$_{CM}$ (keV)\footnotemark[2]&E$_{x}$ (keV)\footnotemark[1] &E$_{x} (keV)\footnotemark[2]$&J$^{\pi}$&$\Gamma_{p}$ (keV)\footnotemark[1]&$\Gamma_{p}$ (keV) \footnotemark[2]\\
\hline
534  $\pm$ 5&0& 0 & 0$^{-}$ & 40 $\pm$ 20 & 25 $\pm$ 5\\
732 $\pm$ 10&193 $\pm$ 6& 198 $\pm$ 10 &1$^{-}$ & $<$ 40 & 70 $\pm$ 5\\
958 $\pm$ 2&424 $\pm$ 5& 425 $\pm$ 2& 2$^{-}$ & 40 $\pm$ 30 & 6 $\pm$ 3\\
\end{tabular}
\end{ruledtabular}
\footnotetext[1]{Recommended values \cite{14}} \footnotetext[2]{This
work.}
\end{table}

The calculation of the $^{15}$O($p$,$\beta^{+}$)$^{16}$O cross
section was made using the properties of $^{16}$F g.s. resonance
measured in the present work and the Breit-Wigner formula for a
single-level resonance \cite{3}:
\begin{eqnarray}
\sigma(E_{p})=\pi \lambdabar^{2}
\frac{2J_{r}+1}{(2J_{i}+1)(2J_{f}+1)}
\frac{\Gamma_{in}\Gamma_{out}}{(E_{p}-E_{R})^{2}+(\frac{\Gamma_{Tot}}{2})^{2}}
\end{eqnarray}
where $\lambdabar$ is the de Broglie wavelength, $J$ are the spins,
and $E_{R}, \Gamma_{Tot}, \Gamma_{in}, \Gamma_{out}$ are the
resonance energy, total width, and partial widths of the incoming
and outgoing channels. In the ($p$,$\beta^{+}$) case, $E_{R} =
E_{g.s.}$ the energy of the g.s. resonance, $\Gamma_{in}$ =
$\Gamma_{p}^{g.s.}$ the proton width, and $\Gamma_{out}$ corresponds
to the $\beta^{+}$-decay partial width. The energy dependance of the
proton width $\Gamma_{p}^{g.s.}(E_{p}$) for the incoming channel was
taken into account by the using the relation:
\begin{eqnarray}
\Gamma_{p}^{g.s.}(E_{p}) =
\Gamma_{p}^{g.s.}(E_{g.s.})\frac{P(E_{p})}{P(E_{g.s.})}
\end{eqnarray}
where $\Gamma_{p}^{g.s.}(E_{g.s.})$ is the proton width at the
resonance energy and $P(E_{p})$ is the penetrability function under
the Coulomb potential barrier. A partial lifetime for
$^{16}$F($\beta^{+}$) of 1 second and a negligible branching ratio
to the $^{15}$O($p$,$\beta^{+}$)$^{12}$C+$\alpha$ final decay
channel were assumed. This assumption is supported by the
$\beta^{-}$-decay properties measured in the mirror nucleus $^{16}$N
\cite{14}. The $\beta^{+}$-decay partial width was taken as a
constant since the energy dependence of the Fermi function is small
due to large Q$_{\beta^{+}}$=15417(8) keV \cite{14}. The calculated
$^{15}$O($p$,$\beta^{+}$)$^{16}$O cross section is shown in Fig. 2
as a function of the center of mass energy. The maximum of the cross
section is observed at the energy of 534 keV corresponding to the
$^{16}$F g.s. resonance. At this energy the $(p,\beta^{+})$ cross
section is very small, about 10$^{-20}$ barns, since $^{16}$F mainly
decays by proton emission, which is $\simeq$ 10$^{20}$ times
stronger than the $\beta^{+}$-decay (since
$\Gamma_{p}^{g.s.}(E_{g.s.})$ = 25 keV and $\Gamma_{out}$ = 0.66
10$^{-18}$ keV).

\begin{figure}[h]
\resizebox{0.40\textwidth}{!}{%
\includegraphics{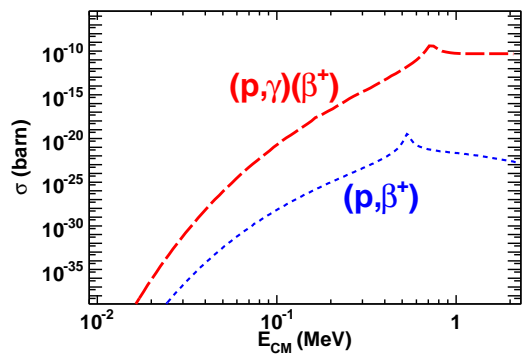}
} \caption{\label{fig:n15}$^{15}$O($p$,$\beta^{+}$)$^{16}$O and
$^{15}$O$(p,\gamma)(\beta^{+})^{16}$O cross sections are shown as a
function of the center of mass energy.}
\end{figure}

\begin{figure}
\resizebox{0.49\textwidth}{!}{%
\includegraphics{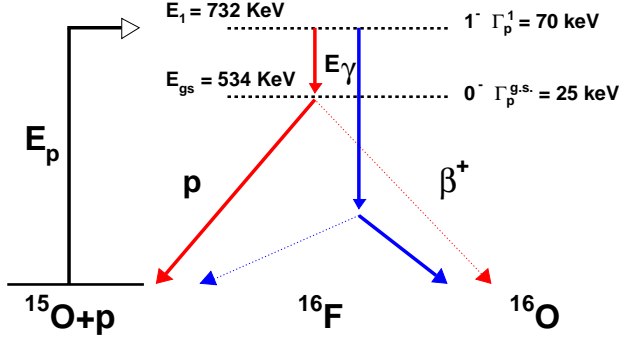}
} \caption{\label{fig:n} Schematic representation of the
$^{15}$O$(p,\gamma)(\beta^{+})^{16}$O reaction (see text). Two cases
are represented. In red, $\gamma$-transitions populate the $^{16}$F
g.s. at the resonance energy. In that case $^{16}$F mainly decays by
proton emission. In blue, high energy $\gamma$-transitions populate
the low energy wing of the g.s. resonance. In that case $^{16}$F
mainly decays by $\beta^{+}$-ray emission.}
\end{figure}

The calculation of the $^{15}$O$(p,\gamma)(\beta^{+})^{16}$O
reaction was performed in a sequential manner, a schematic
representation of this reaction is shown in Fig. 3. A proton capture
reaction to the first excited state of $^{16}$F is considered,
followed by a gamma decay to the g.s. resonance, from which a
$\beta^{+}$-decay branching ratio is taken into account. The cross
section $\sigma_{p\gamma\beta}(E_{p})$ for the
($p$,$\gamma)(\beta^{+}$) reaction at the energy E$_{p}$ is an
integration of the differential cross section over all possible
energies of the gamma transition (since the g.s. has a large width):
\begin{eqnarray}
\sigma_{p\gamma\beta}(E_{p})  = \int
\sigma_{p\gamma}(E_{p},E_{\gamma}) P_{\gamma}(E_{\gamma})
P_{\beta}(E_{p},E_{\gamma}) dE_{\gamma}
\end{eqnarray}
where $\sigma_{p\gamma}(E_{p},E_{\gamma})$ is the cross section to
capture the proton at the energy E$_{p}$ and to emit a $\gamma$-ray
with an energy E$_{\gamma}$, $P_{\gamma}(E_{\gamma})dE_{\gamma}$ is
the strength function \cite{2,15}, that is the probability for the
$\gamma$-ray to have an energy between $E_{\gamma}$ and
$E_{\gamma}$+$dE_{\gamma}$, and $P_{\beta}(E_{p},E_{\gamma})$ is the
branching ratio function for the $^{16}$F nucleus to decay by
$\beta^{+}$-ray emission. The first term
$\sigma_{p\gamma}(E_{p},E_{\gamma})$ is calculated using a
Breit-Wigner formula with the following parameters E$_{1}$,
$\Gamma^{1}_{Tot}(E_{p},E_{\gamma})$, $\Gamma^{1}_{p}(E_{p})$,
$\Gamma_{\gamma}^{1}(E_{\gamma})$ being the energy, total width,
proton width and gamma width for the resonance corresponding to the
first excited state of $^{16}$F. The $\gamma$-ray is emitted from a
1$^{-}$ state to the 0$^{-}$ g.s., which corresponds to a M1
transition, whose energy dependence of the gamma width
$\Gamma_{\gamma}^{1}(E_{\gamma})$ is:
\begin{equation}
\Gamma_{\gamma}^{1}(E_{\gamma}) =
\Gamma_{\gamma}^{1}(E_{1}-E_{g.s.})\{\frac{E_{\gamma}}{E_{1}-E_{g.s.}}\}^{3}
\end{equation}
A gamma lifetime of 1 ps was obtained from the mirror nucleus
\cite{14}, which corresponds to the partial width
$\Gamma_{\gamma}^{1}(E_{1}-E_{g.s.}) = 0.66 10^{-3} eV$. The
strength function of the $^{16}$F g.s. resonance was calculated
assuming a Breit-Wigner parametrization:
\begin{equation}
P_{\gamma}(E_{\gamma})dE_{\gamma}  =
\frac{1}{N}\frac{dE_{\gamma}}{(\Delta
E)^{2}+(\frac{\Gamma^{g.s.}_{Tot}(E_{p}-E{\gamma})}{2})^{2}}
\end{equation}
and the normalization constant is:
\begin{equation} N  = \int\frac{1}{(\Delta
E)^{2}+(\frac{\Gamma^{g.s.}_{Tot}(E_{p}-E{\gamma})}{2})^{2}}dE_{\gamma}
\end{equation}
with $\Delta E = E_{p}-E_{\gamma}-E_{g.s.}$ and
$\Gamma^{g.s.}_{Tot}(E_{p}-E_{\gamma})$$=
\Gamma_{out}+\Gamma_{p}(E_{p}-E_{\gamma})$ is the total width of the
g.s. resonance. The $\beta$ branching ratio is calculated using:
\begin{equation}
P_{\beta}(E_{p},E_{\gamma}) =
\frac{\Gamma_{\beta}}{\Gamma_{\beta}+\Gamma_{p}^{g.s.}(E_{p}-E_{\gamma})}
\end{equation}
Naively, one might have expected to obtain a small cross section for
the $(p,\gamma)(\beta^{+})$ reaction, similar to the $(p,\beta^{+})$
one, since $\gamma$- and $\beta$-widths are much smaller than
proton-widths. Contrary to naive expectations, the
$(p,\gamma)(\beta^{+})$ cross section is about 10$^{10}$ times
larger than the $(p,\beta^{+})$ cross section, as shown in Fig. 2.
The large ratio can be explained in the following way. As it has
been shown previously, there is only one ($p$,$\beta^{+}$) reaction
for 10$^{20}$ ($p,p$) reactions. In the $(p,\gamma)(\beta^{+})$
case, one $\gamma$-ray is emitted for 10$^{8}$ incident protons
(from the ratio of the widths) and about one $\gamma$-transition
 over 10$^{2}$ populates the low energy wing of the
g.s. resonance (less than 50 keV above the proton emission
threshold) where it is almost always followed by a $\beta^{+}$-decay
(P$_{\beta} \simeq 1$). This implies that one incident proton over
10$^{10}$ induces a $(p,\gamma)(\beta^{+})$ reaction, that is a
factor 10$^{10}$ times larger than in the $(p,\beta^{+})$ reaction.

In the following, uncertainties in the calculations and their
evaluated effects on the results are discussed. The position and
width of the low lying $^{16}$F states were measured with a high
precision (see Table 1). The effect of the uncertainties in these
measured parameters results in a change by less than a factor two in
the calculated cross sections. The calculated
$(p,\gamma)(\beta^{+})$ cross section is insensitive to the $^{16}$F
$\beta^{+}$-decay lifetime, as a variation by a factor of 100 causes
the cross section to change by only a factor of 2. The lifetime of
the $\gamma$-transition is a sensitive parameter since the
($p$,$\gamma$)($\beta^{+}$) cross section is almost directly
proportional to this parameter. A value measured in the mirror
nucleus was used, but this assumption works only to within a factor
of 10 \cite{16}. The other excited states in $^{16}$F were also
studied and found to be negligible. The
$^{15}$O($p$,$\gamma$)($p$,$\gamma$)$^{17}$Ne double indirect proton
capture reaction was not taken into account, neither the cross
sections calculated, since it requires an appropriate 3-body
calculation. Moreover, cross sections may change by several effects,
which remain to be evaluated as: non-resonant direct capture
contributions, quantum interferences, continuum couplings \cite{17}.

\begin{figure}[h]
\resizebox{0.48\textwidth}{!}{%
\includegraphics{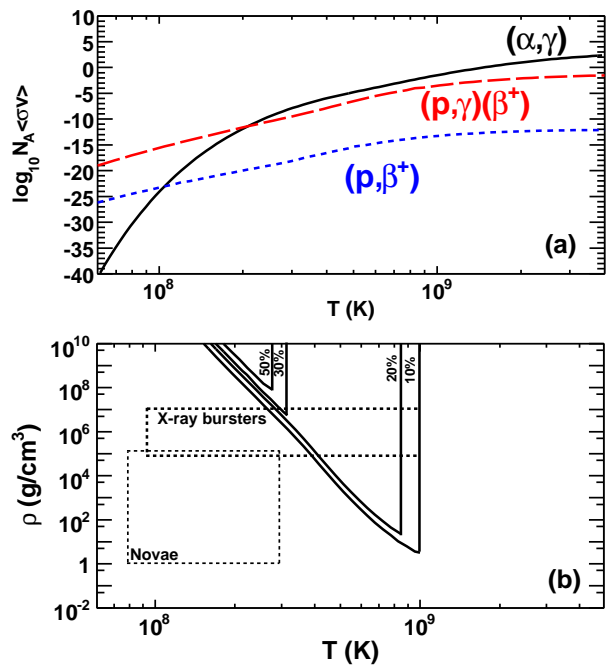}
} \caption{\label{fig:n15}(a) $^{15}$O($p$,$\beta^{+}$)$^{16}$O and
$^{15}$O($p$,$\gamma)(\beta^{+}$)$^{16}$O reactions rates are shown
as a function of the temperature. The
$^{15}$O($\alpha$,$\gamma$)$^{19}$Ne reaction rate is also shown for
comparison. (b) Density versus temperature conditions where the
$^{15}$O($p$,$\gamma)(\beta^{+}$)$^{16}$O reaction represents 10 to
50 \% of the total reaction flux initiated by the $^{15}$O nucleus.}
\end{figure}

At a given temperature T of the gas inside the star, protons exhibit
a Maxwellian distribution and the reaction rates N$_{A}<\sigma v>$
are calculated by integrating numerically the Maxwellian-averaged
cross sections $\sigma(E_{p})$ over all possible proton energies.
The obtained reactions rates are shown in Fig. 4 (a) as a function
of the temperature. The rate of the $(p,\beta^{+})$ reaction is
negligible compared to that of the reaction $(p,\gamma)(\beta^{+})$
for all temperatures. To evaluate the impact of this latter
reaction, it has to be compared with the competing $\beta^{+}$-decay
of $^{15}$O and the $^{15}$O($\alpha$,$\gamma$)$^{19}$Ne alpha
capture reaction. Fig. 4 (b) shows the temperature and density
conditions where the $(p,\gamma)(\beta^{+})$ reaction represents 10
to 50 \% of the total reaction flux initiated by the $^{15}$O
nucleus. Boxes delimit conditions where novae and X-ray bursts can
happen. For the lowest temperatures ($<$ 10$^{8}$ K), the
$(p,\gamma)(\beta^{+})$ reaction requires extreme densities ($>$
10$^{10}$ g cm$^{-3}$) to compete with the
$^{15}$O($\beta^{+}$)-decay. For the highest temperatures ($>$ 1.1
10$^{9}$ K), the $(\alpha,\gamma)$ always dominates. In novae
explosions, $^{15}$O nuclei mainly decay by $\beta^{+}$-ray
emission, the $(p,\gamma)(\beta^{+})$ reaction representing less
than 1 \% of the flux from $^{15}$O. In X-ray bursts, the
$(p,\gamma)(\beta^{+})$ reaction can represent up to 30 \% of the
total flux. Within the uncertainties of the calculations, the
$(p,\gamma)(\beta^{+})$ reaction could be faster than the
$(\alpha,\gamma)$ reaction for temperatures up to 10$^{9}$ K. A more
precise evaluation depends on the $(\alpha,\gamma)$ reaction rate
(not well known) and on the relative abundances in hydrogen and
helium, since one reaction consumes protons and the other alpha
particles. In these extreme conditions, a new cycle of reactions is
operating:
$^{15}$O($p$,$\gamma)(\beta^{+}$)$^{16}$O($p$,$\gamma$)$^{17}$F($p$,$\gamma$)$^{18}$Ne($\beta^{+}$)$^{18}$F($p$,$\alpha$)$^{15}$O.
This new cycle could speed-up the CNO cycle and occur complementary
to breakout reactions. The role of this new proposed cycle of
reactions remains to be studied more carefully under various X-ray
bursts conditions.

In summary, it is shown that unbound nuclei can be involved in
specific reactions that could play a role in astrophysics.
Sequential $(p,\gamma)(\beta^{+})$ reaction, proceeding trough an
intermediate proton-unbound nucleus, was studied for the first time.
The calculated $^{15}$O($p$,$\gamma)(\beta^{+}$)$^{16}$O cross
section is found to be almost 10$^{10}$ times larger than the direct
$^{15}$O($p$,$\beta^{+}$)$^{16}$O reaction cross section. The large
increase is mainly due to a strong feeding of the low energy wing of
the $^{16}$F g.s. resonance, where the subsequent $\beta^{+}$-decay
is favored. The $(p,\gamma)(\beta^{+})$ could act in X-ray bursts,
and would provide a steady burning scenario with a continuous
depletion of $^{15}$O. It is of great importance to study carefully
the effects of this new reaction under various X-ray bursts
conditions, and to demonstrate experimentally the existence of
($p$,$\gamma)(\beta^{+}$) reactions. The cross section of the
$^{15}$O($p$,$\gamma)(\beta^{+}$)$^{16}$O reaction is calculated to
be in the nanobarns range and can be measured using next generation
intense RIB. More generally, several other unbound nuclei as
$^{19}$Na or $^{15}$F could also be involved in this type of
reaction and remain to be studied.  We thank the GANIL crew for
delivering the $^{15}$O beam, M. P\l oszajczak and A. Navin for
stimulating discussions. This work has been supported by the
IN2P3-IFIN-HH Program.


\begin{thebibliography}{}
\bibitem{1}
H.A. Bethe, C.L. Critchfield, Phys. Rev. \textbf{54}, 248 (1938).

\bibitem{2}
C. S. Rolfs and W. S. Rodnay, Cauldrons in the Cosmos, The
University of Chicago Press, Chicago 60637 (1988).

\bibitem{3}
K. Langanke \textit{et al.}, Z. Phys. A \textbf{324}, 147 (1986).

\bibitem{4}
R.K. Wallace and S.E. Woosley, Astrophys. J. Suppl. Ser.
\textbf{45}, 389 (1981).

\bibitem{5}
M. Wiescher, H. Schatz, and A.E. Champagne, Phil. Trans. Roy. Soc.
London A \textbf{356}, 2105 (1998).

\bibitem{6}
H. Schatz \textit{et al.}, Phys. Rev. Lett. \textbf{86}, 3471
(2001).

\bibitem{7}
J. G\"{o}rres \textit{et al.}, Phys. Rev. C \textbf{51},  392
(1995).


\bibitem{8}
P. Bertrand \textit{et al.}, 17th International Conf. on Cyclotrons
and their applications, Tokyo (2004).


\bibitem{9}
R. Anne \textit{et al.}, Nucl. Instr. and Meth. A\textbf{257},215
(1987).

\bibitem{10}
V.Z. Golberg  \textit{et al.}, Phys. At. Nucl. \textbf{60}, 1061
(1997).

\bibitem{11}
F. de Oliveira \textit{et al.}, Eur. Phys. J. A \textbf{24}, 237
(2005).

\bibitem{12}
A.M. Lane and R.G. Thomas, Rev. Mod. Phys. \textbf{30}, 257 (1958).

\bibitem{13}
E. Berthoumieux  \textit{et al.}, Nucl. Instr. and Meth. B
\textbf{136-138}, 55 (1998).

\bibitem{14}
G.Audi \textit{et al.}, Nucl. Phys. A \textbf{729}, 337 (2003); D.R.
Tilley \textit{et al.}, Nucl. Phys. A\textbf{564}, 1 (1993).

\bibitem{15}
A. Messiah, Quantum Mechanics, North-Holland, Amsterdam, Vol. 2
(1962).

\bibitem{16}
F. de Oliveira \textit{et al.}, Phys. Rev. C \textbf{55}, 3149
(1997).


\bibitem{17}
J. Rotureau \textit{et al.}, Phys. Rev. Lett. \textbf{95}, 042503
(2005); R. Chatterjee \textit{et al.}, Nucl. Phys. A \textbf{764},
528 (2006).






\end{thebibliography}

\end{document}